\documentclass[a4paper,11pt]{article}
\usepackage{pos}
\usepackage{graphicx}

\title{One Flavour QCD as an analogue computer for SUSY}
\author[a,b]{Michele Della Morte}
\author[a,b,c]{Benjamin J{\"a}ger}
\author[a,c,d,e]{Francesco Sannino}
\author[a,b]{\newline J.~Tobias Tsang}
\author*[a,b,f]{Felix~P.~G.~Ziegler}

\affiliation[a]{CP3-Origins, University of Southern Denmark, Campusvej 55, 5230 Odense M, Denmark}
\affiliation[b]{Department of Mathematics and Computer Science (IMADA), University of Southern Denmark, Campusvej 55, 5230 Odense M, Denmark}
\affiliation[c]{Danish-IAS, University of Southern Denmark, Campusvej 55, 5230 Odense M, Denmark}
\affiliation[d]{Dipartimento di Fisica “E. Pancini”, Universit`a di Napoli Federico II — INFN sezione di Napoli,
Complesso Universitario di Monte S. Angelo Edificio 6, Via Cintia, 80126 Napoli NA, Italy}
\affiliation[e]{Scuola Superiore Meridionale, Largo S. Marcellino, 10, 80138 Napoli NA, Italy}
\affiliation[f]{School of Physics and Astronomy, The University of Edinburgh, EH9 3FD, Edinburgh, United Kingdom}

\emailAdd{felix.ziegler@ed.ac.uk}

\abstract{We numerically study QCD with a single 
quark flavour on the lattice probing predictions 
from effective field theories that are equivalent to 
minimal super-symmetric Yang-Mills theory in 
the large $N_c$ limit.
The hadronic spectrum including excited states is 
analysed using one gauge coupling and several 
physical volumes and fermion masses. We use the 
LapH method and also compute disconnected 
diagrams.
Lattice simulations with an odd number of Wilson 
fermions give rise to regions of configuration 
space with a negative fermionic weight entailing a 
sign problem. We perform a detailed analysis on 
the spectrum of the Wilson-Dirac operator and 
report on observed cases of a negative fermion 
determinant in our ensembles.}

\FullConference{%
 The 38th International Symposium on Lattice Field Theory, LATTICE2021
  26th-30th July, 2021
  Zoom/Gather@Massachusetts Institute of Technology
}

\begin{document}
\maketitle

\section{Introduction}
In these proceedings we report on our ongoing
efforts to study supersymmetry (SUSY) via lattice computations of the hadronic 
spectrum of $N_f = 1$ QCD. The relation between supersymmetric Yang-Mills theory and QCD with a single quark flavour 
can be seen as follows. The Lagrangian
\begin{equation}
\mathcal{L} = \frac{1}{2g^2} F^a_{\mu \nu}F^a_{\mu \nu} + \overline{\psi} (m_0 + \gamma_{\mu} D_{\mu}) \psi \,,
\label{eq:lark-theory}
\end{equation}
describes a QCD-like 
theory containing a single 
Dirac fermion in the two-index anti-symmetric 
representation of the gauge group $SU(N_c)$ whereas the gluons 
are in the adjoint representation. Note that 
 $\psi^{ij}= \psi^b \ (t^{b})^{ij}\,, \ \ i,j = 1,\ldots,N_c\,, a = 1,
 \ldots,N_c^2-1\,, \ \ b = 1,\ldots,\frac{N_c(N_c-1)}{2}$.
Historically, this theory was studied
in a technicolour extension to QCD at large $N_c$  by Corrigan and Ramond \cite{Corrigan:1979xf}, 
who named the fermion 
field in the two-index anti-symmetric representation a \textit{lark} (merging the words \textit{quark} and \textit{large}). 
Moreover, it was shown that the lark theory (\ref{eq:lark-theory}) 
and $\mathcal{N}=1$ super-Yang-Mills (SYM) theory are equivalent
in the limit $N_c \to \infty$
with regard to the bosonic sector of the spectrum \cite{Armoni:2003gp,Armoni:2004ub}.
Note also that the number of fermionic degrees of freedom scales as $N_c^2$ as $N_c \to \infty$ in both the lark theory and in SYM
signalling their equivalence.\footnote{This can be seen from a simple counting of dimensions. For the lark theory there are 
2 Dirac spinors with 4 components each and the dimension of the anti-symmetric representation scales as 
$N_c^2/2$ for large $N_c$ whereas for SYM there are 
2 Majorana spinors with 2 components each and the adjoint representation scales as $N_c^2$ for large $N_c$.}
For $N_c = 3$ the two-index anti-symmetric representation coincides with 
the conjugate representation, i.e.~a lark is equivalent to 
an anti-quark,
hence (\ref{eq:lark-theory}) describes $N_f = 1$ QCD. 

In SYM the even and odd parity mesons
are degenerate. Deviations from the degeneracy have been studied in the 
lark theory in the large $N_c$ limit in
\cite{Sannino:2003xe} and \cite{Armoni:2005qr}. 
Both works use planar equivalence to predict the lightest 
pseudo-scalar meson to be lighter than the lightest scalar meson.  
In particular, the former work takes into account the explicit SUSY breaking due to the finite fermion mass studying low-energy effective Lagrangians of the lark theory and making use
of exact SUSY results at the effective action level. 
The lark theory by Corrigan and Ramond has led to a plethora of 
applications including investigations of meson scattering \cite{Sannino:2007yp}, a study on (super-)
glue balls in comparison to QCD mesons and glue balls 
\cite{Feo:2004mr},
investigations of the conformal window \cite{Sannino:2004qp, Dietrich:2006cm} as well as works in
phenomenology \cite{Hong:2004td,Dietrich:2005jn}.

In this work we simulate $N_f = 1$ QCD on the lattice to probe the mentioned prediction and to study relics of SYM for $N_c = 3$.
This also comes with the advantage of lower simulation costs compared to direct lattice simulations of SYM, the latter being hard because massless fermions need to be handled. However, it should be emphasised that $N_f=1$ QCD should be regarded 
if at all as a proxy for SUSY. 

A few years back a lattice study probing the planar equivalence prediction was presented at the Lattice conference by the M{\"u}nster group and collaborators \cite{Farchioni:2008na}.
Our work at hand can be regarded as an update in an advanced setup using tree-level
$O(a)$-improved Wilson fermions. In addition, we also extract excited states of the mesonic spectrum.

Unlike in the continuum formulation where the fermion 
determinant is guaranteed to be positive 
%for a positive fermion mass 
in the lattice formulation with Wilson fermions there are regions
of configuration space with a negative fermionic weight.
This gives rise to a sign problem. We present a detailed analysis on the sign of the fermion determinant on our gauge field ensembles.
This account is organised as follows: In Section \ref{sec:setup} we summarise our lattice setup, followed by the sign problem analysis in 
Section \ref{sec:sign}. In Section \ref{sec:hadrons} we show the results on the hadronic spectrum and we conclude this work in Section \ref{sec:concl}.

\section{Lattice setup}
\label{sec:setup}

The lattice action considered in this work contains the tree-level 
Symanzik improved gauge  action and the tree-level $O(a)$-improved Wilson 
clover fermion action ($c_{SW} = 1$).
We restrict ourselves to a single gauge coupling corresponding to 
$\beta = 4.5$. 
To simulate the single quark flavour the RHMC algorithm 
\cite{Kennedy:1998cu, Clark:2006fx} is used.
$N_f = 1$ QCD comes with two main challenges. (i) The scale setting cannot be carried out in the usual way as in e.g.~$N_f=2+1$ QCD by using an experimentally known (low-energy) quantity such as a hadron mass. (ii) chiral symmetry is absent which 
excludes comparisons with chiral perturbation theory. 
Addressing challenge (i), we obtain an 
approximation to the scale by setting the lattice spacing 
using the Wilson flow in the pure gauge
theory following 
\cite{Luscher:2010iy} which results in $a \approx 0.06 \ \mathrm{fm}$.
Regarding challenge (ii) it is noteworthy that even in the absence of 
chiral symmetry it is possible to guarantee (at least approximately) 
a well-defined extrapolation to zero quark mass. To that end the 
mass of the lightest pseudo-scalar meson, called the 
\textit{fake pion}, 
is measured in the partially quenched extension of the single 
flavour theory obtained by adding an additional valence quark, see 
\cite{Farchioni:2008na} for details and references.
Simulating $N_f=1$ QCD 
amounts to navigating in unknown territory in parameter space
for the various mentioned reasons. 
To our knowledge neither chiral perturbation theory nor the method put forward in 
\cite{Luscher:1985dn,Luscher:1986pf} for estimating finite volume effects have been worked out
for the lark theory in general. However, we consider these effects to be sub-leading at the level of precision we are interested in.
Therefore we have produced and analysed gauge field
ensembles for several physical volumes 
$L/a \in \{12,16,20,24,32\}, T/a = 64$
and hopping parameters\footnote{The hopping parameter is defined as $\kappa=\frac{1}{2 (4 + m_0)}\,$.} $\kappa \in \{0.1350,0.1370,0.1390,0.1400,0.1405,0.1410\}$.
All configurations for this project have been generated with 
the \texttt{openQCD} software package \cite{openQCD}. 

\section{Sign problem}
\label{sec:sign}

As mentioned above in a setup of a single flavour of Wilson fermions at finite lattice spacing there exist regions 
of configuration space on which the fermion determinant 
is negative. This can also occur in multi-flavour QCD and has been subject to a recent study in $N_f=2+1$ flavour QCD \cite{Mohler:2020txx}. Since we are generating our gauge field
ensembles with respect to the sign quenched fermionic weight 
we need to monitor the sign of the fermion determinant on configurations we measure observables on and in case a negative determinant is detected it has
to be accounted for by reweighting. Since a direct computation 
of the sign of the fermion determinant is numerically too expensive, we infer it indirectly from the low-lying
eigenvalue spectrum of the Wilson Dirac operator $D$. 
$\gamma_5$-hermiticity of $D$ guarantees the eigenvalues to 
come either in complex conjugated pairs or to be real. 
This entails that only real eigenvalues can produce a sign change
of the determinant. Hence, on a given configuration 
it remains to check if there is an odd number of negative real eigenvalues
of $D$.

 In practice it is more convenient to consider the Hermitian matrix $Q:=\gamma_5 D$.
Because $\det(Q) = \det(D)$ and since a zero eigenvalue of $D$ 
is also a zero eigenvalue of $Q$ we can thus reduce the sign computation to analysing the behaviour of the low eigenvalues 
$\lambda_i(m_0)$ of $Q$ as a function of the bare mass $m_0$. 
Monitoring the sign and counting the zero crossings 
of the low eigenvalues of $Q$
when varying $m_0$, the change in the number 
of negative real eigenvalues of $D$ can be inferred.

\begin{figure}[t]
\includegraphics[width=.5\textwidth]{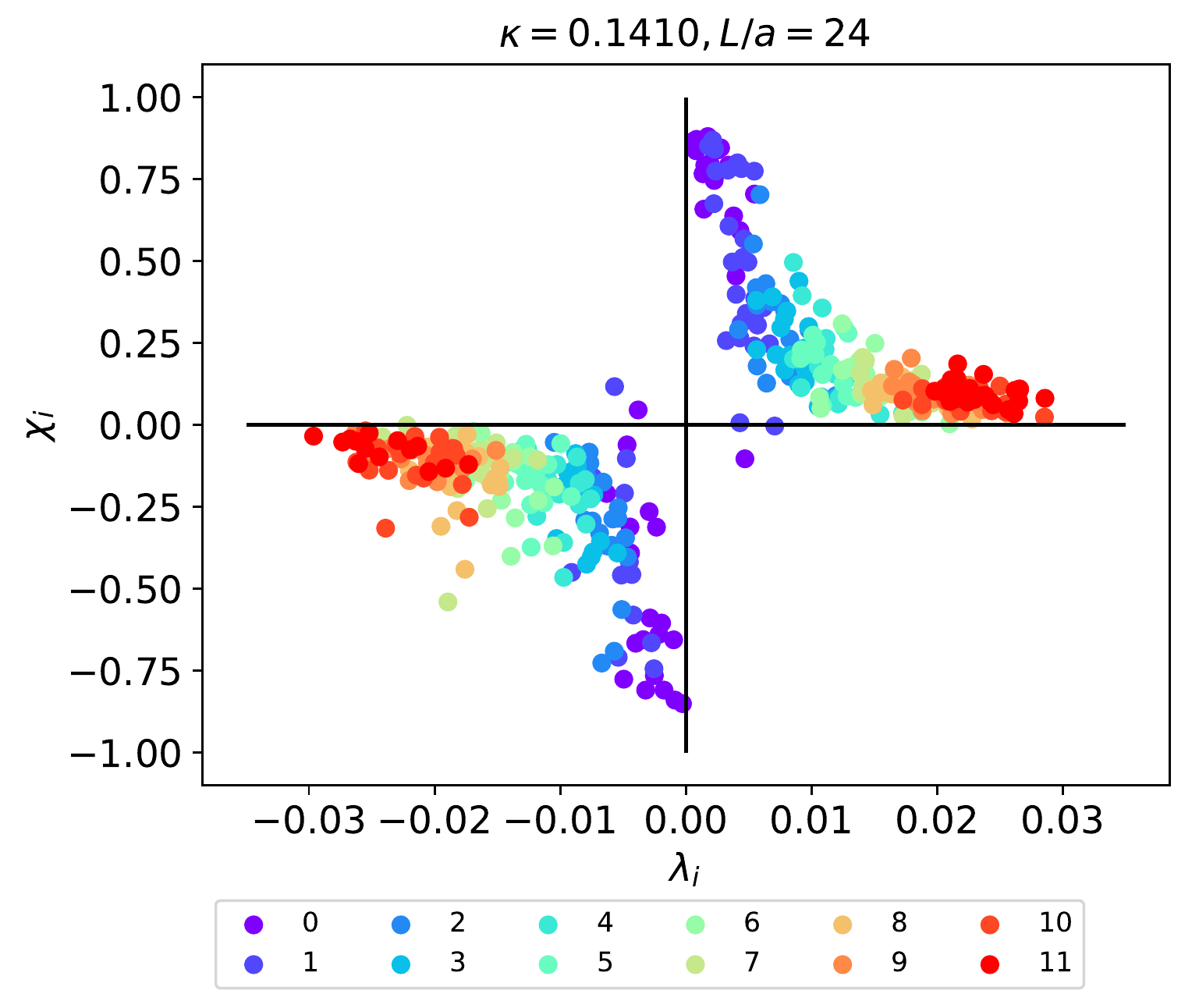}
\includegraphics[width=.5\textwidth]{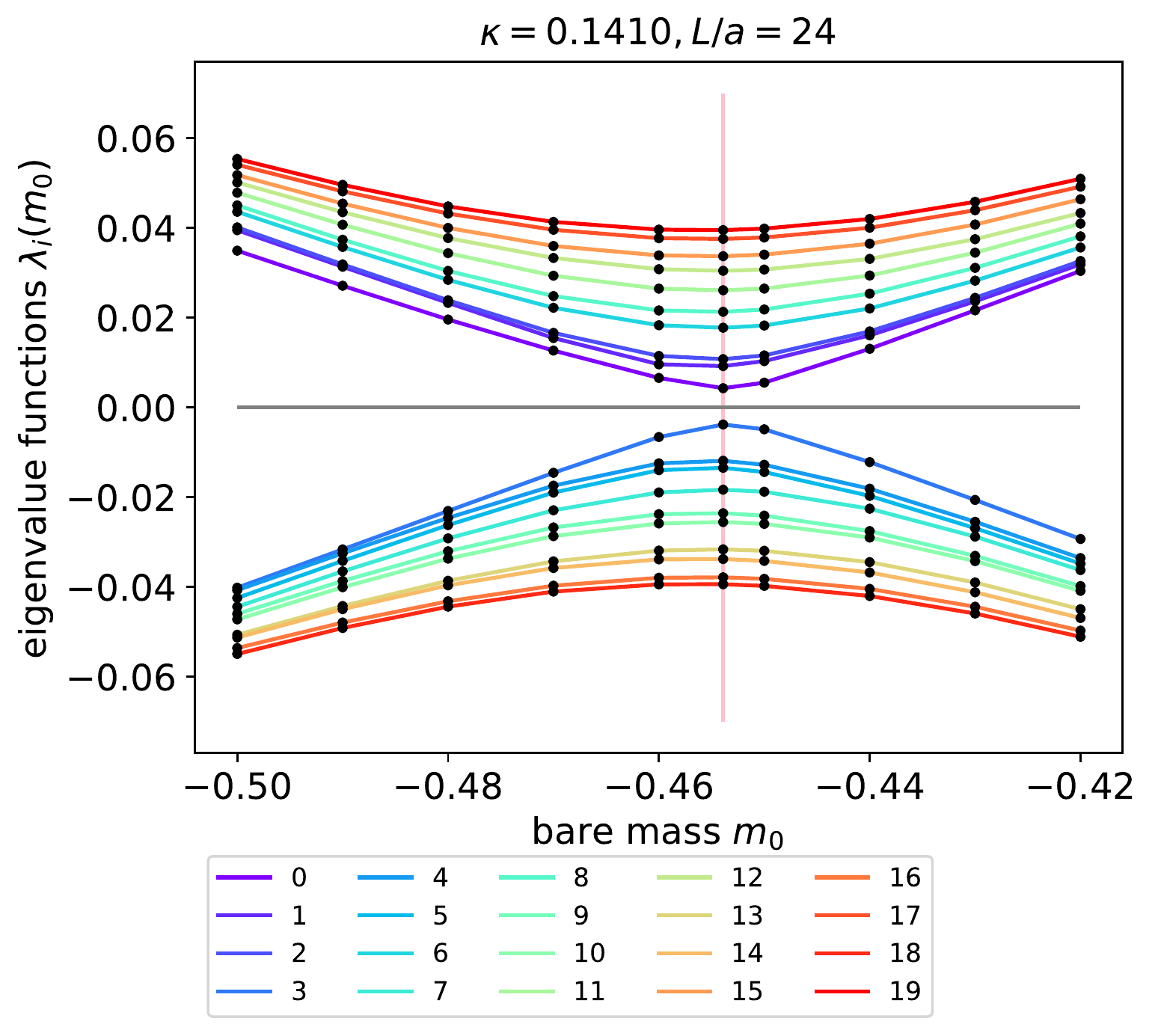}
\caption{\emph{Left:} Showcase of stage (i) of the determinant sign 
analysis. 
The 12 lowest eigenvalues and their corresponding 
chiralities of $Q$ 
are computed on a subset of ca.~40 
configuration from the $L/a = 24$ ensemble at $\kappa = 
0.1410$. It is visible that there are a few points in the second and 
fourth quadrant signalling that the corresponding 
eigenvalues at this bare mass move towards zero as the mass is 
infinitesimally increased. These configurations are further checked 
in stage (ii) of the analysis with regard to whether and how many 
sign changes of the eigenvalue functions appear.\\
\emph{Right:} Showcase of stage 2 of the determinant sign analysis. 
Displayed is the partially quenched computation of the lowest 20 eigenvalues of $Q$
%$(D^{\dagger} D)^{1/2}$  
on a fixed gauge configuration in the $(L/a = 24, \kappa=0.1410)$
ensemble for a range of values for the bare mass around the 
simulation mass $m_0^* \approx -0.453$ indicated by the pink vertical line. This configuration is safe in the sense
that the fermion determinant is positive as there are no sign 
changes of the eigenvalue functions $\lambda_i(m_0)$.} 
\label{fig:ev-stage-1-2}
\end{figure}

Technically, we proceed in two stages: (i) For a given configuration 
simulated with a bare mass $m_0^*$ we compute the lowest lying
eigenpairs 
$(\lambda_i, \phi_i)$ of $Q$. 
 In addition, the chirality $\chi_i = d\lambda_i/ d m_0|_{m_0=m_0^*} = 
(\phi_i,\gamma_5 \phi_i)$ being the slope of the eigenvalue function~\cite{Mohler:2020txx, Akemann:2010em} is computed. 
Hence from the 
sign of the chirality we can infer if a given eigenvalue moves towards or away from zero as $m_0$ is increased infinitesimally.
A showcase for stage (i) of the analysis is displayed in the left panel of Figure \ref{fig:ev-stage-1-2}.

\begin{figure}[t]
\begin{center}
\includegraphics[width=.5\textwidth]{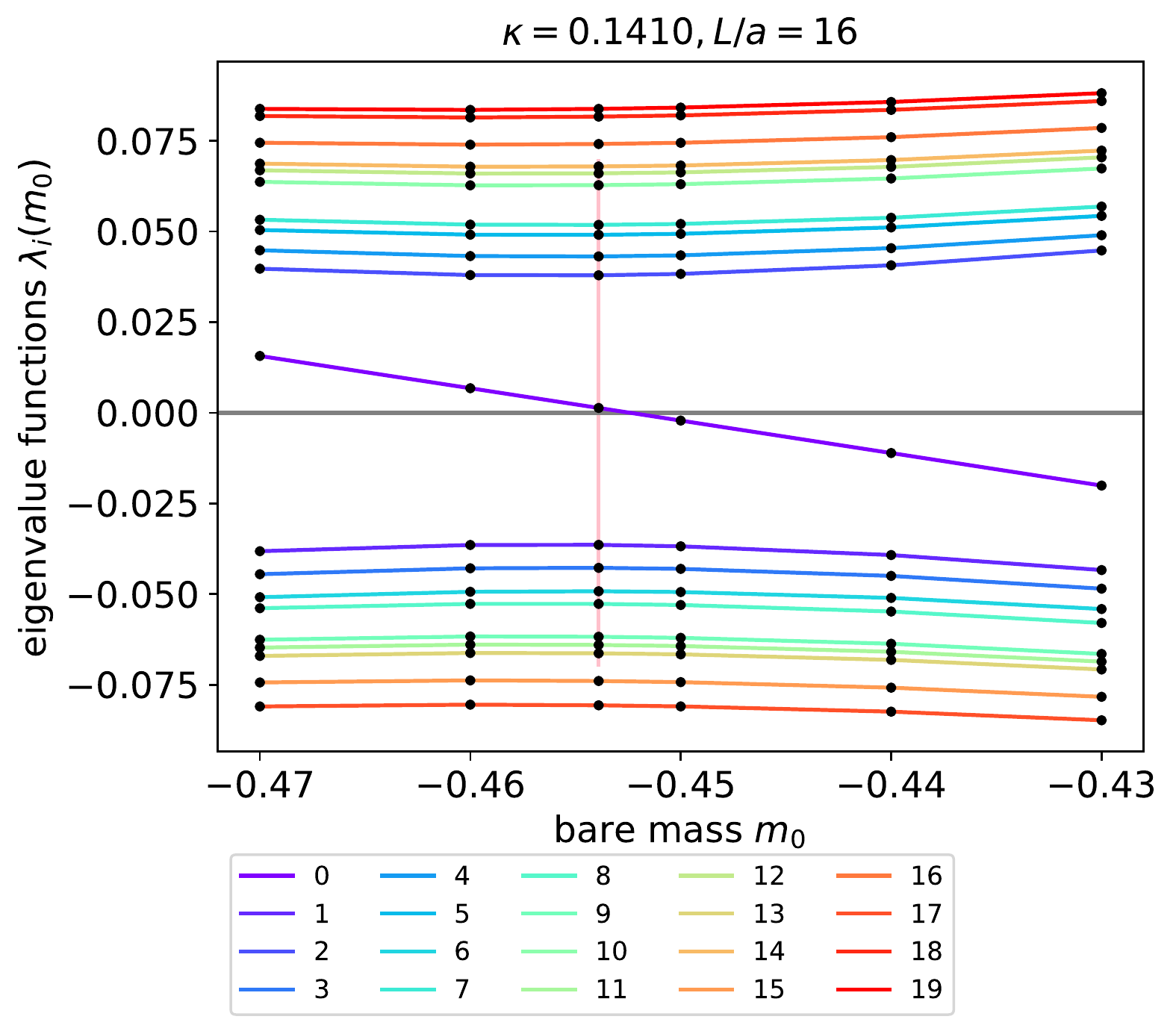}
\end{center}
\caption{Showcase of stage 2 of the tracking analysis where
the occurrence of a negative fermionic determinant can be 
deduced from a single sign change in one of the low eigenvalues
of $Q$ above the simulation mass $m_0^*$. From this we can also
conclude that $D$ must have an odd number of negative real eigenvalues at the simulation mass.}
\label{fig:ev-neg-det}
\end{figure}

(ii) Configurations giving rise to low eigenvalues of $Q$ that might cross zero are further analysed using the
tracking method presented in \cite{Mohler:2020txx}. 
On a given configuration we measure the lowest $N_{\mathrm{ev}}$ eigenpairs $(\lambda_i, \phi_i)$ of $Q$ for 
several bare masses $m_0$ around the simulation mass. 
The measurements were carried out using  
the \texttt{PRIMME} package \cite{PRIMME, svds_software} combined with
\texttt{openQCD}.  Assuming that $\mathrm{span}(\{\phi_i\})$ changes slowly and
continuously as $m_0$ is varied in steps of $\Delta m_0$ allows to extract the
eigenvalue function $\lambda_i(m_0)$ by matching the basis vectors $\phi_i(m_0)$
and $\phi_j(m_0 + \Delta m_0)$ with respect to their overlap.  The right panel
of Figure \ref{fig:ev-stage-1-2} displays a showcase for the tracking method
where the conclusion is that the fermion determinant is positive on the analysed
configuration.  We measured the sign of the fermion determinant only for $\kappa
= 0.1410$ on the $L/a = 16,24$ ensembles.  For both volumes the occurrence of
$\det(D) < 0$ is less than $1 \%$ of the configurations which were taken into
account for measuring correlators and masses shown in Section
\ref{sec:hadrons}. From this we conclude that the sign problem is mild for the
parameters and volumes investigated in this work.  Figure \ref{fig:ev-neg-det}
shows one of the rare cases we found where the sign of the determinant is
negative. Here a single eigenvalue function changes sign once at a bare mass larger than the simulation mass. 
Note that in the limit of  $m_0 \gg 0$ $\det(D)$ is 
positive and so are all real eigenvalues of $D$. 
Decreasing $m_0$ to the simulation mass $m_0^*$ and finding an odd number of zeros of the $\lambda_i(m_0)$ implies that $\det(D) < 0$ at $m_0^*$.

In the ensembles generated at smaller values of $\kappa$ the sign problem is even milder as an increasing mass shifts the eigenvalues of $D$ to larger values decreasing the likelihood of a negative determinant to occur.
\section{Hadron spectroscopy}
\label{sec:hadrons}
We are interested in the massless limit of the mesonic spectrum of $N_f=1$
  QCD. We are using the LapH
  method~\cite{Morningstar:2011ka,HadronSpectrum:2009krc} which provides a
  suitable framework to include quark-disconnected pieces, which are vital to
  correctly extract the spectrum. It furthermore allows to cheaply construct
  different operators $\hat{O}_i$ inducing the same quantum numbers and hence
  the same spectrum but differing in the approach to the ground state. We are
  interested in the masses of the pseudo-scalar (P), scalar (S), and vector (V)
  mesons. In principle a scalar glueball can be present, so in addition to
  operators of the type $\bar{q} \Gamma q$ we also use a purely gluonic operator
  that is expected to predominantly couple to the glueball (G). We extract the
  spectrum by performing simultaneous correlated three-exponential fits to
  several correlation functions using the ansatz
\begin{equation}
C_{ij}(t) = \sum_n \langle 0 |\hat O_i| n \rangle \langle n | \hat O_j^{\dagger} | 0 \rangle  e^{-m_n t}\,.
\end{equation}

\begin{figure}[t]
\includegraphics[width=.5\textwidth]{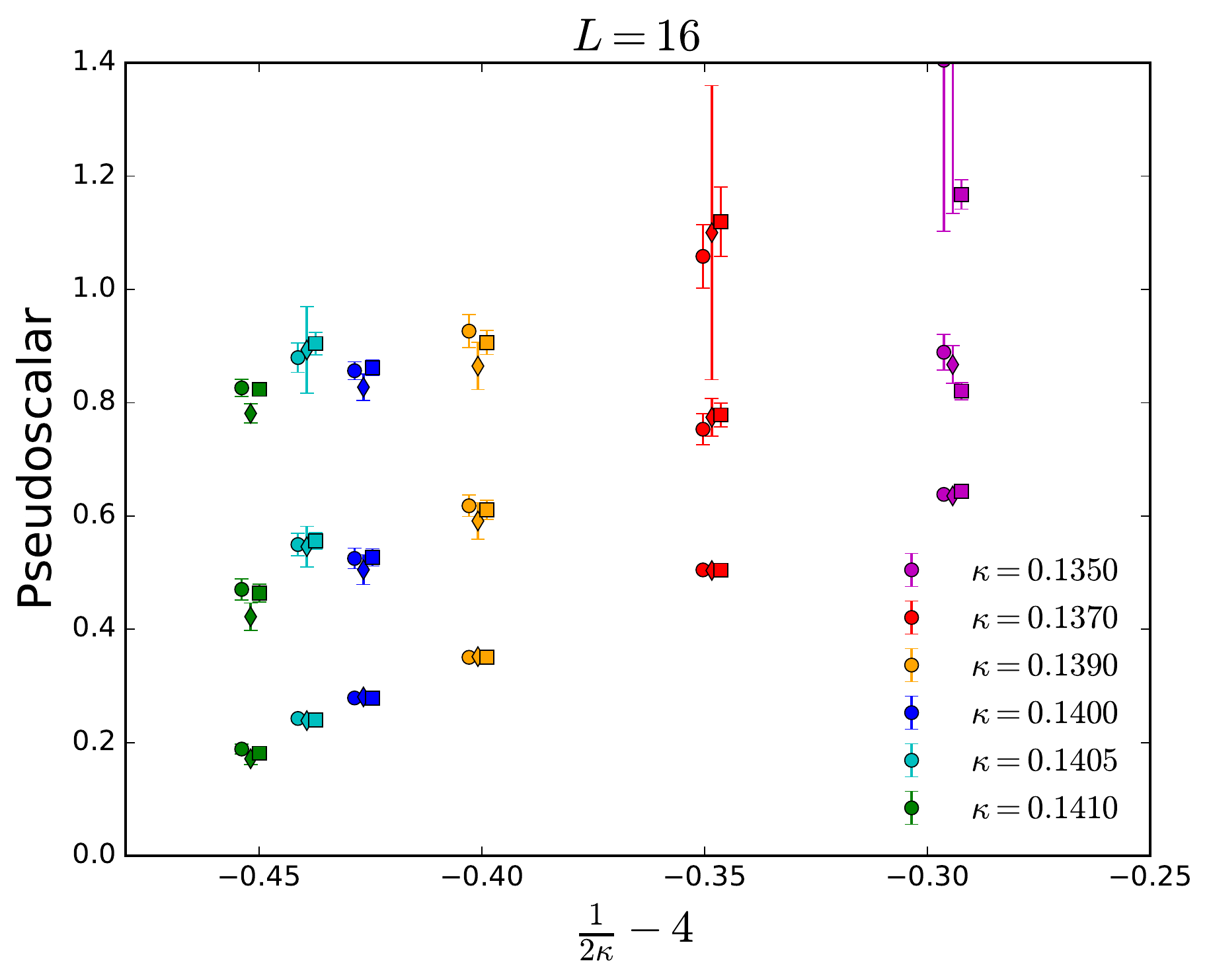}~
\includegraphics[width=.5\textwidth]{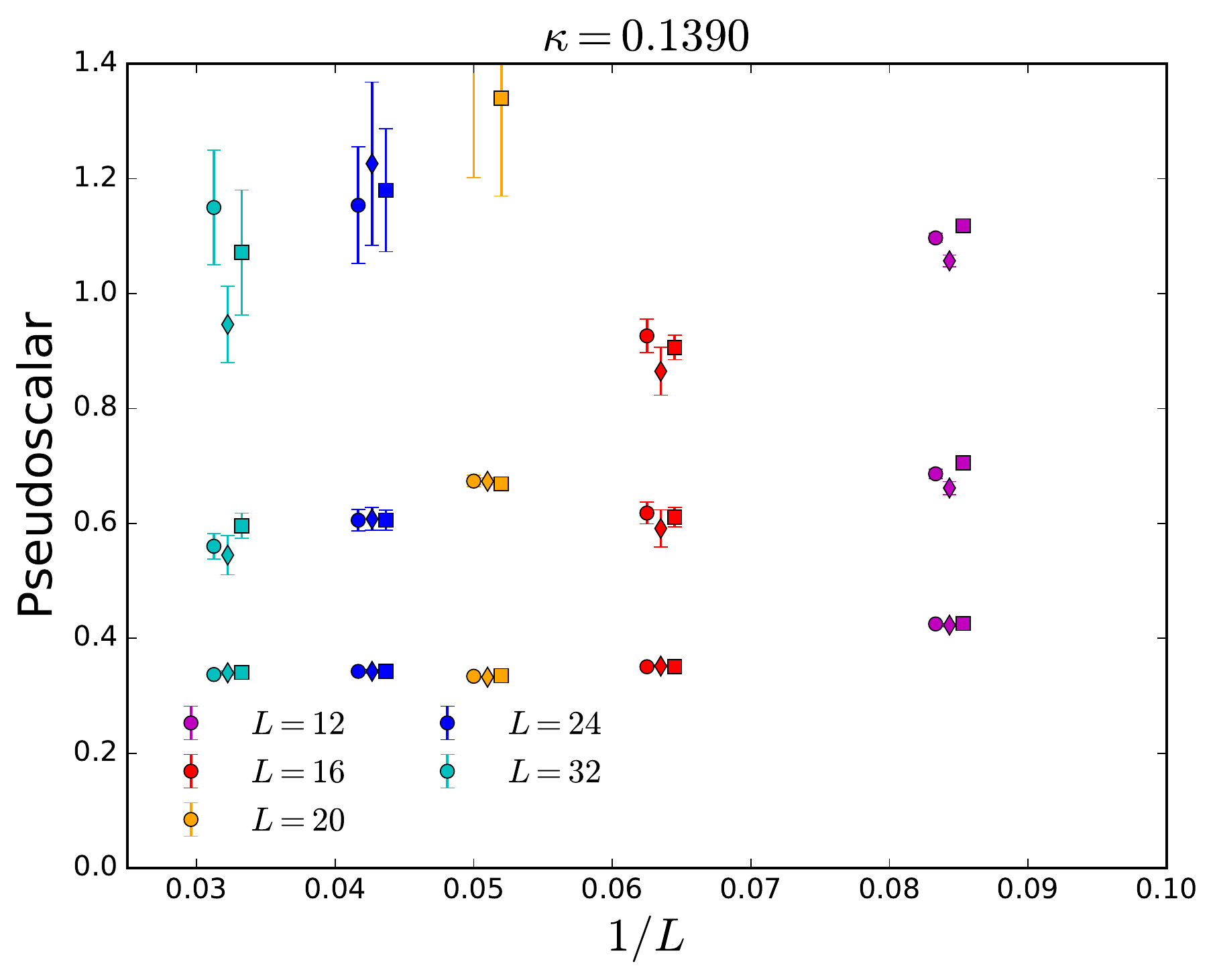}
\caption{\emph{Left:} Spectrum  
of the pseudo-scalar meson for the various bare masses $m_0$ 
at fixed volume $L/a = 16$. \emph{Right:}
Spatial volume dependence of the spectrum 
of the pseudo-scalar meson  
at fixed $\kappa = 0.1390$.}
\label{fig:p-kappa-vol}
\end{figure}

In Figure~\ref{fig:p-kappa-vol} the result for the spectrum of the pseudo-scalar
meson is shown. As a measure of stability we perform each for different
  subsets of the operators in the basis, shown by the different symbols (circles, squares, diamonds).  In
the left panel the bare quark mass dependence is shown at fixed volume
($L/a=16$). As expected the ground state is strongly mass dependent. In
  the right panel we investigate the volume dependence of the spectrum 
at fixed bare quark mass corresponding to $\kappa = 0.1390$. At this value of
$\kappa$, we find the ground state to be volume independent, but with sizable
finite size effects being displayed for the smallest volume.

\begin{figure}[t]
\includegraphics[width=.5\textwidth]{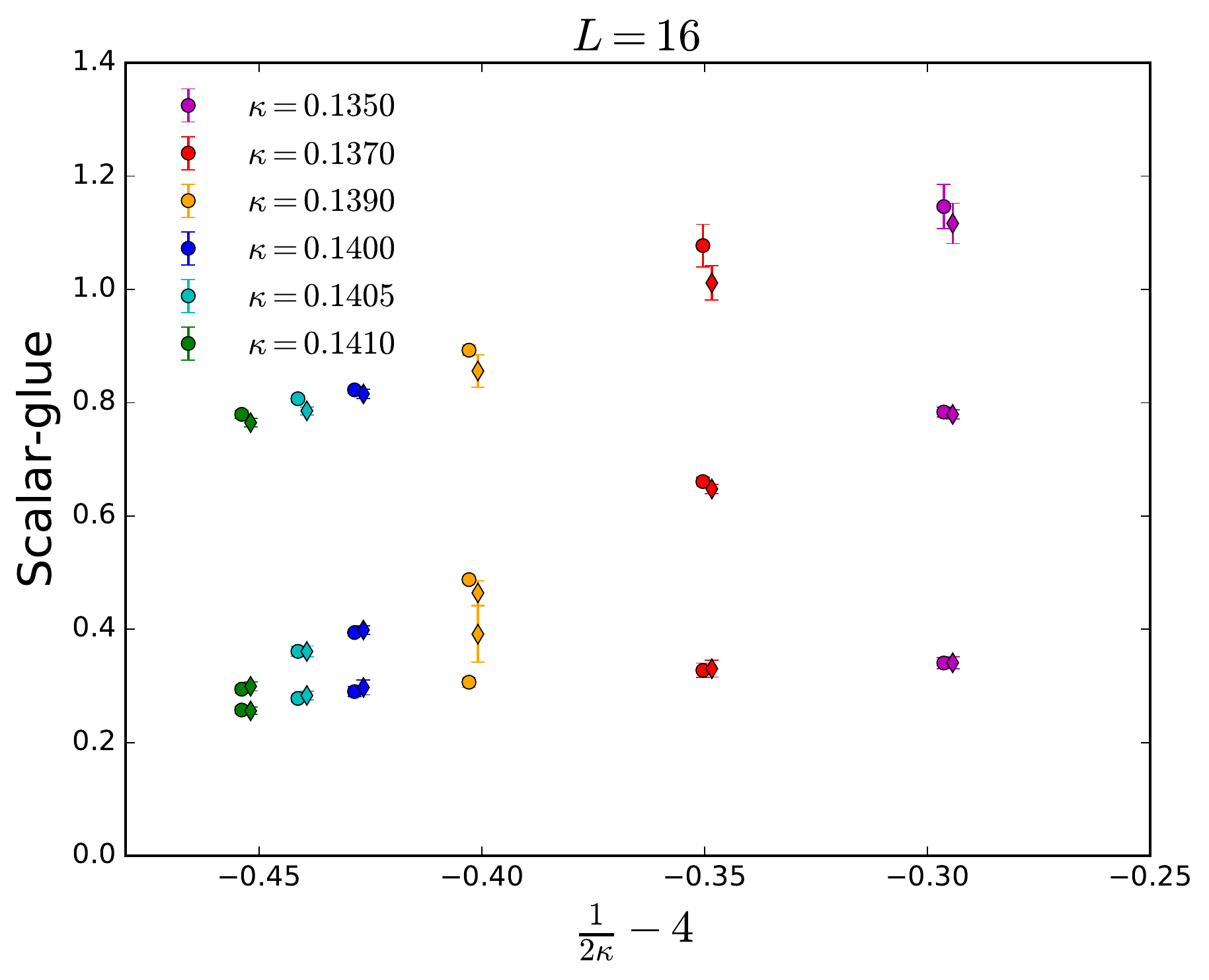}~
\includegraphics[width=.5\textwidth]{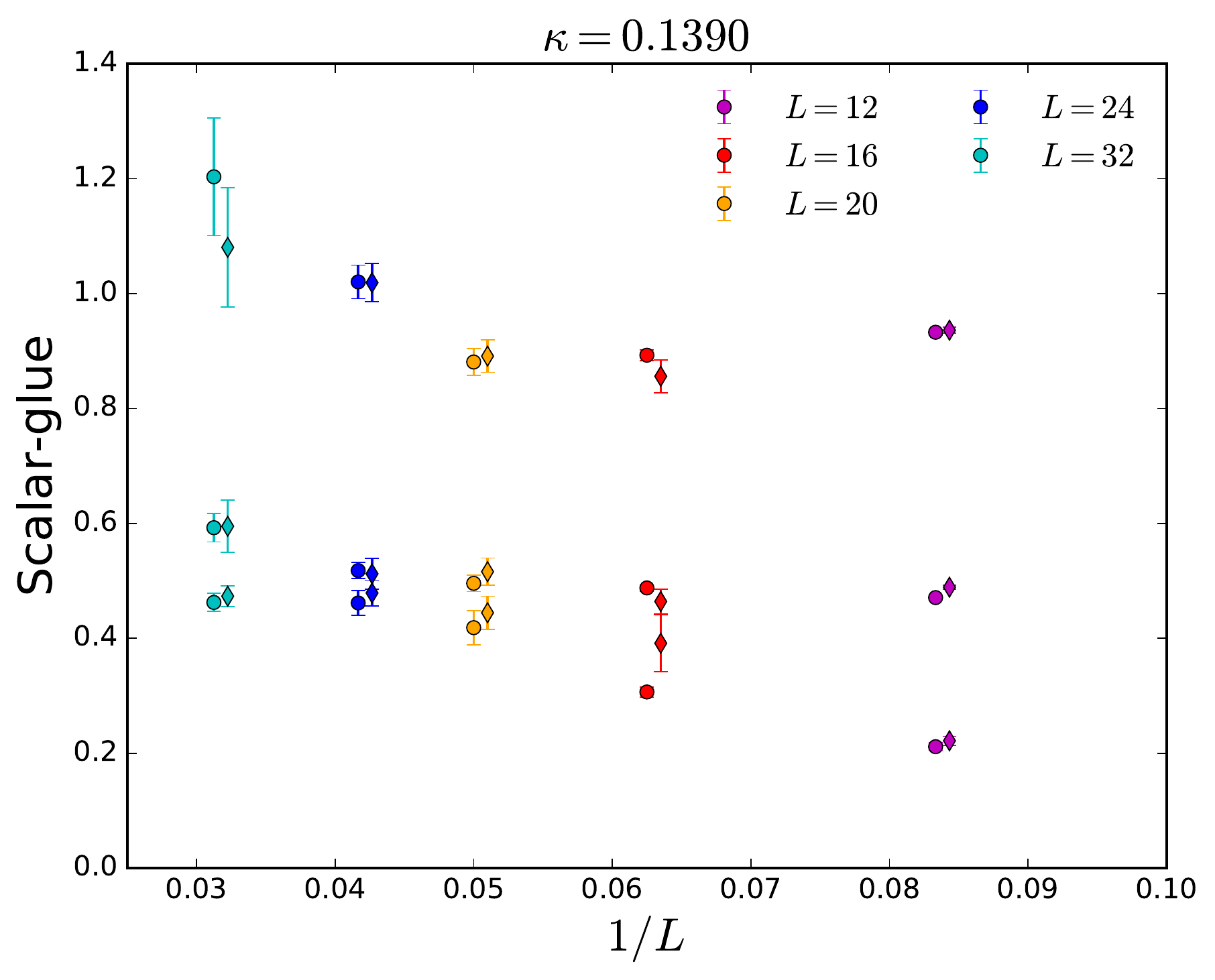}
\caption{\emph{Left}: Spectrum  
of the scalar meson for the various bare masses $m_0$ 
at fixed volume $L/a = 16$. \emph{Right}: 
Spatial volume dependence of the spectrum 
of the scalar meson  
at fixed $\kappa = 0.139$.}
\label{fig:s-kappa-vol}
\end{figure}

\begin{figure}[t]
\includegraphics[width=.5\textwidth]{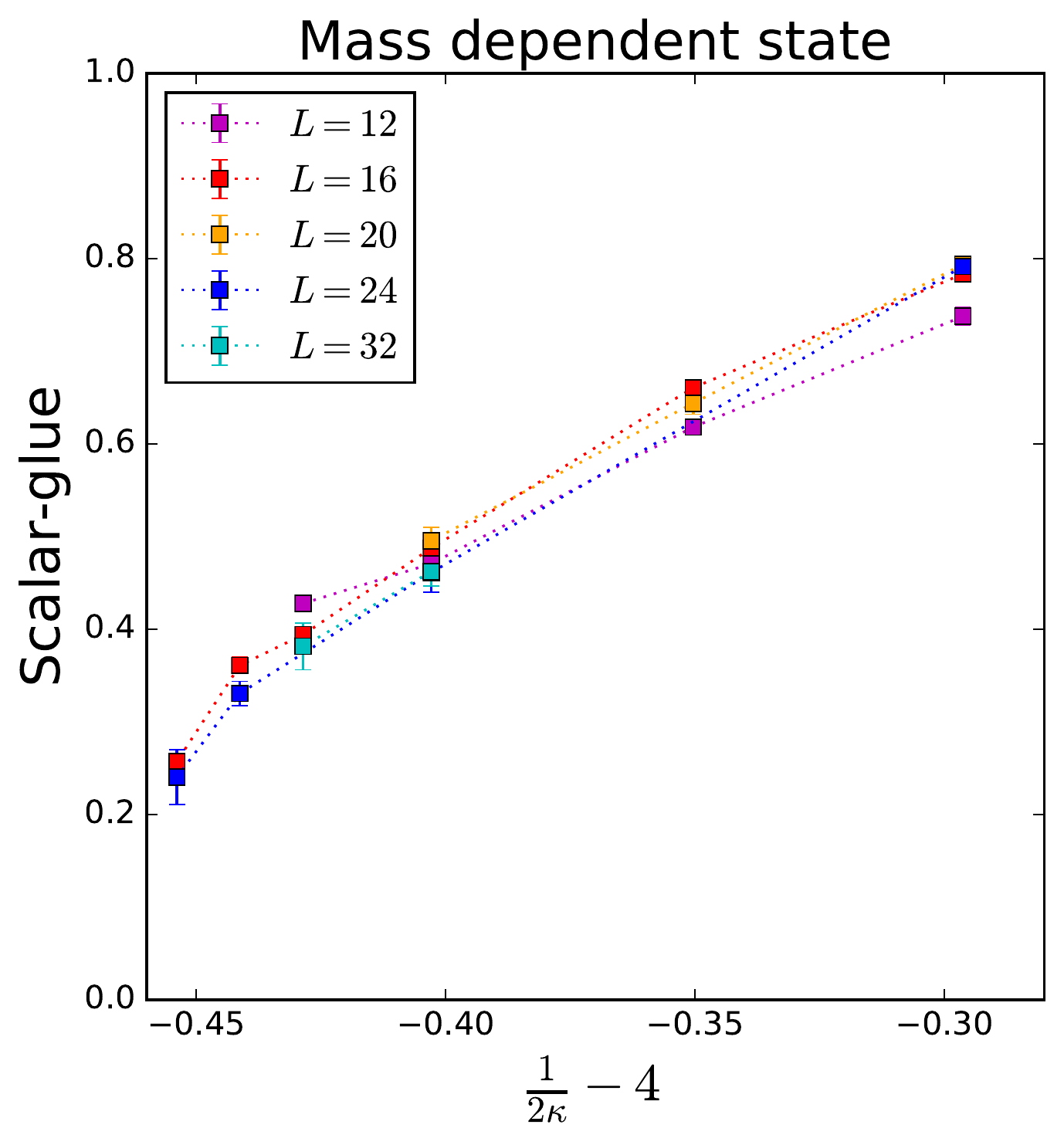}~
\includegraphics[width=.5\textwidth]{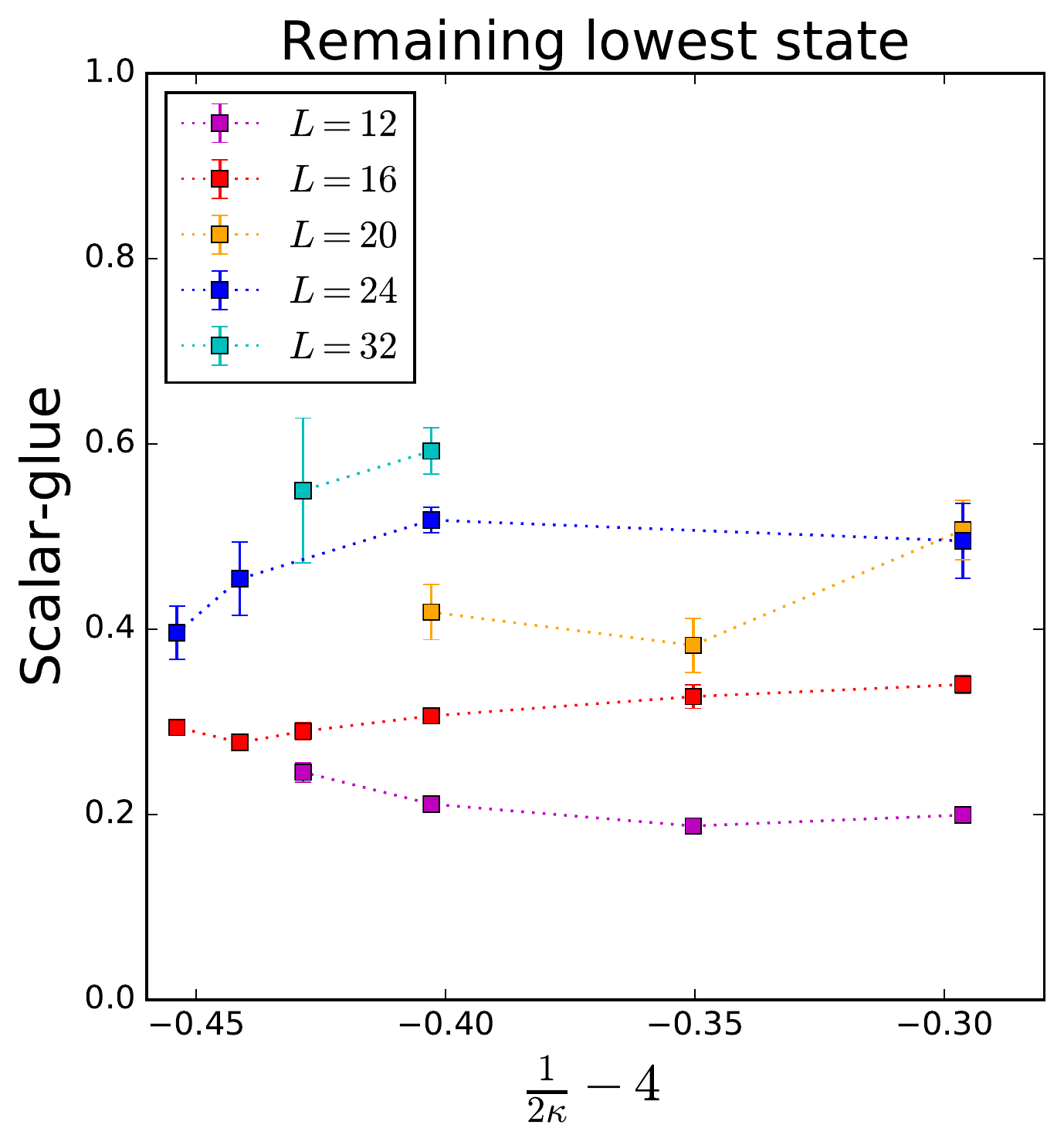}
\caption{Categorisation of the spectrum of the scalar meson into a
mass-dependent (left) and a remaining lowest state (right).
In the right-hand side plot the strong volume dependence is 
inconsistent with what is expected for a glueball state. 
A possible interpretation is that this state is a finite volume torelon state.}
\label{fig:s-mass-other}
\end{figure}

For the scalar-glueball channel the extracted spectrum looks qualitatively
different as shown in Figure \ref{fig:s-kappa-vol}. The left panel shows the
mass dependence at the same fixed volume.  Looking at the ground and first
excited state, the lightest observed state is mass-insensitive at fixed volume
whilst the first excited state displays a clear mass dependence.  In the right
panel of the same figure we again investigate the volume dependence of these
states at fixed mass.  For this value of $\kappa$, there appears to be one
volume dependent state and one volume independent state with a cross-over
somewhere between $L/a = 16$ and $L/a= 20$. We repeat this comparison for each
volume and in each case identify the mass-sensitive state with a scalar
meson. This identification is guided by monitoring the behaviour of the overlap
factors, for example whether the correlation function purely built from the
gluonic operators couples more strongly to the ground or the first excited
state. The resulting data points are shown in the left hand panel of
Figure~\ref{fig:s-mass-other}.  From the right hand panel of
Figure~\ref{fig:s-mass-other} we find that the remaining lowest state is
strongly volume dependent, getting heavier as the volume increases.  This is
inconsistent with the expected behaviour for a glueball state, so we identify
this state to be a finite volume state possibly resulting from flux tubes around
the periodic lattice, also called a torelon state
\cite{Michael:1985ne}. However, in the study at hand we have not further
investigated the character of these states and whether they are torelon
states. In principle this could be done by computing correlators of spatial
Wilson loops.

\begin{figure}[t]
\begin{center}
\includegraphics[width=.5\textwidth]{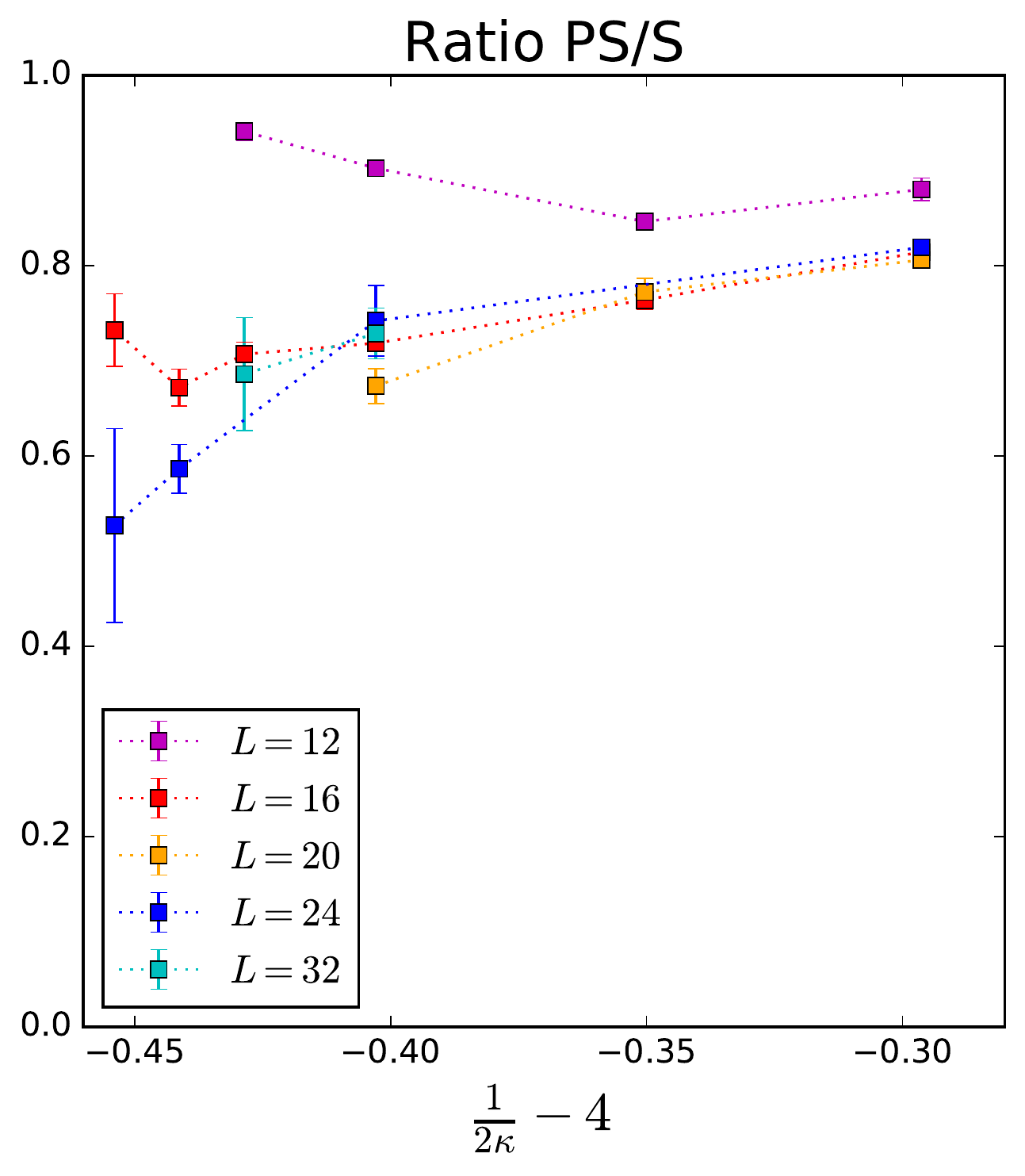}
\end{center}
\caption{Ratio of the pseudo-scalar to scalar meson mass
as a function of the various bare masses $m_0$.}
\label{fig:ratio}
\end{figure}

Figure \ref{fig:ratio} shows the mass ratio of the (mass-dependent)
pseudo-scalar to scalar meson. Disregarding the smallest volume which appears to show strong finite size effects the results for the 
different volumes are in agreement in the range
between large and intermediate 
masses. 
Our data confirms that the pseudo-scalar meson is lighter than
the scalar one. 
For a robust comparison with the low-energy effective theory 
prediction from \cite{Sannino:2003xe} we need to perform an extrapolation to vanishing quark mass. This is subject to ongoing work and we are currently producing more statistics at the largest volumes and hopping parameters.

\section{Conclusion and perspectives}
\label{sec:concl}
We have presented an analysis on the hadronic spectrum of $N_f=1$ QCD in
the mesonic sector.  By including excited states we have extracted the mass
dependency of the scalar and pseudo-scalar state. The next step of performing
the extrapolation to zero quark mass is subject to ongoing work.  From this,
comparisons with the predictions from low-energy effective theories for the
deviation from the even-odd parity degeneracy can be made.  Loosely speaking,
this will also show what remnant SUSY is contained in the $N_c = 3$ lark theory.
Moreover, we have shown that the sign problem due to the use of Wilson fermions
is mild but must be monitored. We emphasise the relevance of this aspect also in
multi-flavour QCD simulations.

Future work will be devoted to investigating the lark theory on 
the lattice for larger number of colours $N_c > 3$. To that end 
we are working on code development and the use of GPU accelerators.
\newline
\newline
\textit{Acknowledgements:} We thank John Bulava for his
contributions in the early stages of this project. We thank Felipe Attanasio and Antonio Rago 
for discussions. The
project leading to this application has received funding from the European
Union's Horizon 2020 research and innovation programme under the Marie
Sk{\l}odowska-Curie grant agreement No 894103. M.D.M. and J.T.T. are partially supported by DFF Research project 1. Grant n. 8021-00122B.
The computing resources for this work were provided by the U. of Southern Denmark and DeiC Interactive HPC (UCloud).
\newpage
{
\bibliography{references}
\bibliographystyle{apsrev4-1ow}
}

\end{document}